\documentclass[11pt]{iopart}
\usepackage{graphicx}
\usepackage{epsf,amsfonts,amssymb,amsbsy,wrapfig,cite}
\begin{document}
\title[Radial geodesics as a microscopic origin of black hole entropy]
{Radial geodesics as a microscopic origin of black hole entropy.\\
I: Confined under the Schwarzschild horizon}
\author{V.V.Kiselev*\dag}
\address{*
\ Russian State Research Center ``Institute for
High Energy
Physics'', 
Pobeda 1, Protvino, Moscow Region, 142281, Russia}
\address{\dag\ 
Moscow
Institute of Physics and Technology, Institutskii per. 9,
Dolgoprudnyi Moscow Region, 141700, Russia}
\ead{kiselev@th1.ihep.su}
\begin{abstract}
Causal radial geodesics with a positive interval in the
Schwarzschild metric include a subset of trajectories completely
confined under a horizon, which compose a thermal statistical
ensemble with the Hawking-Gibbons temperature. The Bekenstein--Hawking entropy  
is given by an action at corresponding geodesics of particles with
a summed mass equal to that of black hole in the limit of large
mass.
\end{abstract}


\section{Introduction}

At present we can observe many faces of black hole entropy first
motivated by Bekenstein \cite{Bekenstein} and Hawking
\cite{Hawking}. In addition to the thermodynamical determination
by an information loss and radiation of black body, it was
calculated through an action of gravitational field itself
\cite{Hawking2,Carlip,Solovev}. In a sense, a dual representation
in terms of quantum fields seized by a black hole, was introduced
in \cite{FrolovNovikov}. The most impressive evaluations were
obtained in conformal field theories \cite{Carlip2,Strominger} and
in superstrings \cite{StromingerVafa+}, wherein the entropy was
derived by counting microstates. Other aspects were revealed by
Loop Quantum Gravity \cite{Ashtekar}. Recent reviews on the
subject, containing a complete list of references, can be found in
\cite{Primer,Damour,Fursaev,Das}.

In present paper, a new insight is offered: we maximally extend a
physical space of causal geodesics of Schwarzschild solution. A
notion of `causal' implies a time-like value of interval on the
geodesics. Then we explore the action on geodesics completely
confined under a horizon in order to calculate a partition
function, which actually results in the Bekenstein--Hawking
relation between the area of horizon with the entropy for a
particular case of Schwarzschild solution. A thermal equilibrium
of confined geodesics requires a thermal quantization of levels.
The space-time of geodesics confined under the horizon is conic
\cite{Susskind}. Thus, we present a primitive microscopic picture
for an internal structure of black hole.

In section 2 we classify causal radial geodesics and thermally
quantize those of confined under the horizon. Section 3 is devoted
to the evaluation of entropy, while discussion, short comparison
with other approaches and conclusion are collected in section 4.

\section{Radial geodesics}
In the Schwarzschild metric
\begin{equation}\label{schwarz}
    {\rm d}s^2=\left(1-\frac{r_g}{r}\right){\rm
    d}t^2-\frac{1}{\displaystyle 1-\frac{r_g}{r}}{\rm d}r^2-r^2[{\rm
    d}\theta^2+\sin^2\theta\,{\rm d}\phi^2]
\end{equation}
the radial geodesics depend on a single integral of motion $A$
found by Hilbert, so that
\begin{equation}\label{geos}
    v^2\equiv\left(\frac{{\rm d}r}{{\rm d}t}\right)^2 =
    \left\{1-A\left(1-\frac{r_g}{r}\right)\right\}\left(1-\frac{r_g}{r}\right)^2,
\end{equation}
where $r_g=2 M G/c^2$ is the Schwarzschild radius, giving the well
kown horizon.

Indeed, a particle motion in the metric of (\ref{schwarz}) is
determined by the Hamilton--Jacobi equation
\begin{equation}\label{1}
    g^{\mu\nu}\;{\partial_{\mu} S_{HJ}}\,{\partial_{\nu} S_{HJ}} - m^{2} =
    0,
\end{equation}
where $m$ denotes the particle mass. Following the general
framework, we write down the solution in the form, which
incorporates the single integral of the radial motion in the
spherically symmetric static gravitational field,
\begin{equation}\label{2}
    S_{HJ} = -{\cal E}\, t+{\cal S}_{HJ}(r),
\end{equation}
where $\cal E$ is the conserved energy. Then, from (\ref{1}) we
deduce
\begin{equation}\label{3}
    \left(\frac{\partial {\cal S}_{HJ}}{\partial r}\right)^2 =
    \frac{1}{g^2_{tt}(r)}\,{\cal E}^2-\frac{m^2}{g_{tt}(r)},
    \quad\mbox{at}\quad g_{tt}(r)=1-\frac{r_g}{r},
\end{equation}
which results in
\begin{equation}\label{4}
    {\cal S}_{HJ}(r) = \int \limits_{r_0}^{r(t)} \textrm{d}r\;
    \frac{1}{g_{tt}(r)}
    \sqrt{{\cal E}^2-V^2(r)},
\end{equation}
where $V^2$ is an analogue of potential,
\begin{equation*}
V^2(r) = g_{tt}(r)\,m^2.
\end{equation*}
The trajectory is implicitly determined by equation
\begin{eqnarray}
  \frac{\partial S_{HJ}}{\partial {\cal E}} &=& \textsf{const} = -t
  +\int\limits_{r_0}^{r(t)} \textrm{d}r\; \frac{1}{g_{tt}(r)}
    \frac{\cal E}{\sqrt{{\cal E}^2-V^2(r)}}. \label{p1}
\end{eqnarray}
Taking the derivative of (\ref{p1}) with respect to the
time\footnote{As usual $\partial_t f(t) =\dot f$.}, we get
\begin{eqnarray}
  1 &=& \dot r\; \frac{\cal E}{g_{tt}(r)\,
  \sqrt{{\cal E}^2-V^2(r)}}, 
\end{eqnarray}
resulting in (\ref{geos}) at $A=m^2/{\cal E}^2$.

At the trajectory the interval takes the following form:
\begin{equation}\label{ds}
    {\rm d}s^2 =\frac{A}{\displaystyle 1-A+A\frac{r_g}{r}}\,{\rm d}r^2,
\end{equation}
so that introducing
\begin{equation}\label{rc}
    r_c=-\frac{A}{1-A}\,r_g,
\end{equation}
we get
\begin{equation}\label{dsrc}
    {\rm d}s^2=\frac{r_c}{r_g}\,\frac{r}{r_c-r}\,{\rm d}r^2.
\end{equation}
In addition to light cones, causal geodesics are given by
trajectories with a positive interval
\begin{equation}\label{>0}
    {\rm d}s^2 > 0\quad \Leftrightarrow\quad \{r < r_c,\;r_c>0\}\cup\{r_c<0\}.
\end{equation}
We subdivide the region of admissible $r_c$-values to three ranges
or classes,
\begin{equation}
\begin{array}{rccc}
  CI:   & r_c<0 &\Leftrightarrow & 0< A \leqslant 1, \\[2mm]
  CII:  & r_g < r_c < +\infty &\Leftrightarrow & 1< A < +\infty,
  \\[2mm]
  CIII: & 0 < r_c< r_g & \Leftrightarrow & -\infty< A < 0.
\end{array}
\end{equation}
At all of above trajectories a proper time is finite at a path
under the Schwarzschild horizon, i.e. at $r<r_g$.

In range $CI$, the probe particle can reach a spatial infinity
$r\to +\infty$, where its velocity is given by $v^2_\infty =1-A$.
In $CII$ the particle \textit{always} has a path outside the
horizon, but the motion is restricted by $r<r_c$ at $r_c>r_g$.
Finally, in $CIII$ the geodesics are \textit{completely confined}
under the horizon. Note, that $r_c=r_g$ is given by $A=\pm\infty$,
which, hence, can be identified, so that all ranges are
continuously connected. This fact implies that external forces
(for instance, collisions of particles) changing the gravitational
integral of motion $A$ (or $r_c$), can transport or convert a
trajectory from one range to another\footnote{In the quantum
mechanics, the interaction should have matrix elements
non-diagonal in $A$.}.

The range\, $CII$ can be completely described in isotropic Kruskal
coordinates $\{\bar u,\bar v\}$ defined by
\begin{equation}\label{i1}
\left\{\begin{array}{l}
 u=t-r_*,\\
 v=t+r_*,\end{array}
 \right.\qquad r_*=r+r_g\ln\left[\frac{r}{r_g}-1\right],
\end{equation}
and
\begin{equation}\label{i2}
    \left\{\begin{array}{l}
\vspace*{-10mm}\\
\displaystyle\bar u=-2r_g\,
e^{\displaystyle-\frac{u}{2r_g}}
,\\[2mm]
\displaystyle\bar v=+2r_g\, e^{\displaystyle+\frac{v}{2r_g}}
,
\end{array}
 \right.
\end{equation}
wherein the metric takes the form
\begin{equation}\label{is}
    {\rm d}s^2 = \frac{r_g}{r}\cdot e^{\displaystyle-\frac{r}{r_g}}\cdot{\rm
    d}\bar u\,{\rm d}\bar v-r^2({\rm d}\theta^2+\sin^2\theta{\rm
    d}\phi^2),
\end{equation}
so that the singularity at the horizon $r=r_g$ is absent, but it
actually appears at $r=0$ as well as $r=\infty$, where the
isotropic metric coefficient respectively becomes equal to
infinity and zero (why $CI$ cannot be completely included in the
Kruskal coordinates).

The inverse transition is given by
\begin{equation}\label{id1}
    \left\{\begin{array}{l}
\displaystyle t= r_g\ln\left[-\frac{\bar v}{\bar u}\right],\\[4mm]
\displaystyle r_* =r_g\ln\left[-\frac{\bar u\cdot\bar
v}{4r_g^2}\right],
\end{array}
 \right.
\end{equation}
so that the singular point is
\begin{equation}\label{id2}
    r=0 \quad\Leftrightarrow\quad \frac{\bar u\cdot\bar
v}{4r_g^2}=1,
\end{equation}
while the horizon corresponds to axes of $\{\bar u,\bar v\}$:
\begin{equation}\label{id3}
    r=r_g\quad\Leftrightarrow\quad \bar v=0 \,\cup\, \bar u=0.
\end{equation}

\begin{figure}[th]
\begin{center}
\setlength{\unitlength}{1.3mm}
\begin{picture}(70,72)
 \put(70,62){$\bar v$}
 \put(0,62){$\bar u$}
 \put(30,53){$r=0$}
 \put(30,15){$r=0$}
 \put(53,48){$H_+$}
 \put(53,20){$H_-$}
 \put(5,33){\fbox{II}}
 \put(65,33){\fbox{I}}
 \put(31,42){\fbox{III}}
 \put(31.5,25){\fbox{IV}}
 \put(0,0){\includegraphics[width=70\unitlength]{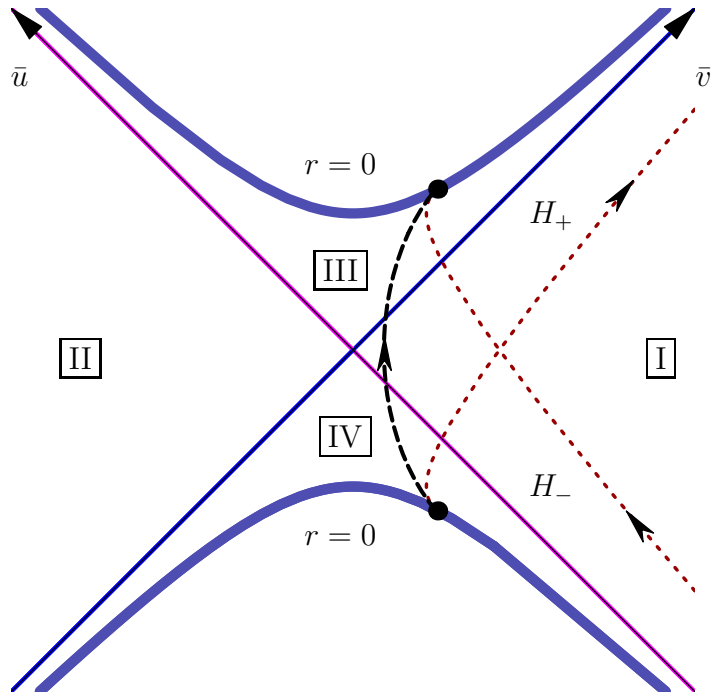}}
\end{picture}
\end{center}
\caption{\small Curves of singularity $r=0$ restrict the maximal
admissible physical region in the Kruskal variables $\{\bar u,\bar
v\}$. The region of external observer in the Schwarzschild metric
I is limited  by the horizon of future $H_+$ and of past $H_-$:
for such the observer the future of events in the ``black hole''
III and past of events in the ``white hole'' IV are not available.
An observer in region II is not causally related with an observer
in I, and regions IV and III represent ``black and white holes''
for him respectively (in II the time $t$ `flows from top to
bottom'). A dashed curve is geodesic of $CII$: any two points at
top and bottom curves of singularity can be connected by a
geodesic curve in $CII$. A short-dashed curve shows a part of
geodesic from the family of $CI$.} \label{pic1}
\end{figure}

The differentials are given by
\begin{equation}\label{dif}
\left\{\begin{array}{l}
 \displaystyle {\rm d}t=r_g\frac{{\rm d}\bar v}{\bar v}-r_g\frac{{\rm
 d}\bar u}{\bar u},\\[4mm]
 \displaystyle {\rm d}r_*=r_g\frac{{\rm d}\bar v}{\bar v}+r_g\frac{{\rm
 d}\bar u}{\bar u},
\end{array}
 \right.
 \qquad\qquad {\rm d}r_*=\frac{{\rm d}r}{1-\frac{r_g}{r}},
\end{equation}
which have a singularity at the horizon.

Light cones for the radial motion in metric (\ref{is}) are
determined by
\begin{equation}\label{ilight}
    {\rm d}\bar u =0\quad\cup\quad {\rm d}\bar v=0.
\end{equation}

The causal structure of space-time in the Kruskal coordinates is
shown in \mbox{Fig. \ref{pic1},} where a characteristic example
for geodesics of $CII$ is also depicted by a dashed curve, while a
short-dashed curve represents a part of geodesic from the family
of $CI$.

At $A>0$ we can easily integrate out (\ref{geos}) at $r\to r_g$ to
get a dependence of radial coordinate on the time near the
horizon, so that
\begin{equation}\label{r-t}
    t-\xi\,r_* \to\mbox{const.}\quad \mbox{at}\quad r\to r_g,
\end{equation}
where $\xi=\mbox{sign}[\dot r]$ is a velocity signature.
Therefore, a body falling to the horizon ($\xi=-1$) has a finite
value of $\bar v$, while a body coming out of the horizon
($\xi=+1$) has a finite value of $\bar u$ as shown in
Fig.\ref{pic1}. Further, differentials (\ref{dif}) give
\begin{equation}\label{deriv}
    \frac{{\rm d}\bar u}{{\rm d}\bar v} = -\frac{\bar u}{\bar v}\,
    \frac{1-\dot r_*}{1+\dot r_*},
\end{equation}
so that in vicinity of singularity
\begin{equation}\label{near}
    \frac{{\rm d}\bar u}{{\rm d}\bar v}\approx \frac{\bar u}{\bar
    v}\,\left(1-\xi\,\sqrt{\frac{r}{A r_g}}\right),\qquad r\to 0,
\end{equation}
and it has the following limit independent of $A>0$:
\begin{equation}\label{cross}
\frac{{\rm d}\bar u}{{\rm d}\bar v}= \frac{4r_g^2}{\bar
v^2},\qquad r=0.
\end{equation}
These characteristics have been taken into account in Fig.
\ref{pic1}. Note, that the isotropic Kruskal coordinates are not
complete, since the trajectories coming to infinity should be
described in Schwarzschild coordinates at $r> R_0>r_g$ (otherwise
the nullification of metric coefficient will appear as an
asymptotic reaching the light cone by the trajectory\footnote{An
ordinary compactifying transformation can help, see
\cite{BirrelDavies}.}).

\begin{figure}[th]
\centerline{\includegraphics[width=7cm]{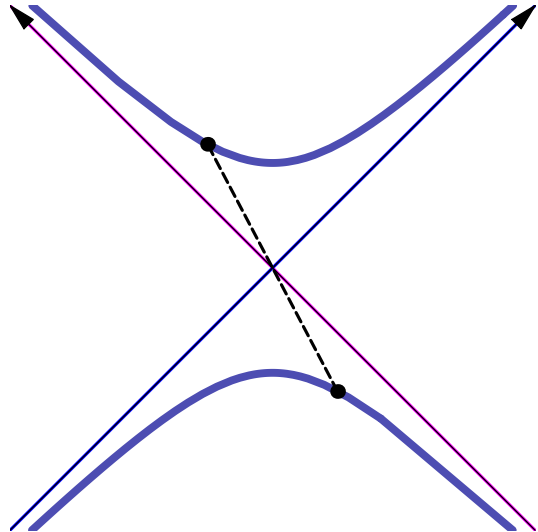}}
  \caption{A common geodesic for $CII$ and $CIII$.
  Labels of axes are the same as in Fig. \ref{pic1}.}\label{pic1a}
\end{figure}

Class $CIII$ has a common element with class $CII$: the geodesic
lines $\bar u=\varkappa^2\bar v$, $0<\varkappa^2<+\infty$. An
example of common geodesics is depicted in Fig. \ref{pic1a}. Note,
that by (\ref{cross}) all geodesics in $CI$ and $CII$, crossing
the singularity at the same point of $\{\bar u,\bar v\}$ plane,
have the same tangent line, which is the extremal geodesic line at
$A\to \infty$, as shown in Fig. \ref{pic1a}.

However, the space-time of geodesics in the range $CIII$ cannot be
described in terms of real $\{t,r\}$ or $\{\bar u,\bar v\}$, since
a positive interval at geodesics is expressed as
\begin{equation*}
    {\rm d}s^2 =A\left(1-\frac{r_g}{r}\right)^2\,{\rm d}t^2 > 0,
\end{equation*}
and at $A<0$ it gives ${\rm d}t^2<0$, i.e. the imaginary
differential of time.

Since we deal with causal solutions of Einstein equations for the
metric with the given asymptotic behavior at $r\to \infty$, we
have to appropriately adjust a coordinate system suitable for the
description of solutions. In general, there are two options:
\mbox{\textit{i}) solution} should be constrained to avoid
problems with some choices of coordinates, i.e. the continuation
of solution is forbidden; \mbox{\textit{ii}) the solution} is
continuable, while coordinate systems have constrained regions of
applicability. We follow the second option, which composes the
paradigm of general relativity: the gravitational dynamics
determines the space-time structure, i.e. causal geodesics compose
the physical space-time\footnote{We have to specially emphasize
that \textit{i)} fixing the metric signature by $\{+,-,-,-\}$
\textsl{at any point of physical space-time} as a postulate of
general relativity forbids the class $CIII$ of geodesics inside
the black hole, \textit{ii)} excluding the singularities on global
maps (by contrast with local ones) forbids the crossing of
horizons, i.e., for instance, it restricts the Schwarzschild
solution by $r>r_g$, and the continuation of solution as the black
hole is forbidden, too. Requiring the absence of singularities on
a global coordinate map of space-time (in a global reference
frame) is essentially equivalent to the postulate on the identity
of general relativity and theory of tensor field in the Minkowski
space-time. The introduction of singularities (horizons) in the
global sense, i.e. the description of space-time by regular local
maps, means, as we see, the refusal of the field-theoretical
interpretation of general relativity and recognizing its pure
geometric nature (with following problems of introducing an
$S$-matrix in the quantum theory and so on). In the latter case we
have to consider the signature of metric, which is asymptotically
given as a boundary condition for the geometric equations of
gravity, as the dynamical geometric quantity. Then the class of
geodesics $CIII$ under study is not only admissible, but it is
conceptually necessary for the consistency and completeness of
theory. Thus, in addition to the mathematical formalism of general
relativity everybody should decide to himself (herself) which
dynamical principles are admissible, acceptable or not.}.

I shall postulate, that a full physical space-time of test
particles is constructed by the following procedure of thought
experiment: Imagine that a test particle is a point-like
spacecraft equipped by an engine. By definition, the engine can
change the integral of motion (continuously, if the system is
classical, or discretely, if the system is quantum), by preserving
the causality: the interval of massive particle has to be positive
under the action of engine\footnote{A mathematical question is a
somehow smoothness or topology. I will leave it free, since a ball
situated on a plane has a single common point with the plane, not
a two-dimensional transition, so that a connection is not smooth
nor regular, and it is topologically peculiar. Analogously, a cone
situated on a plane also has a common one-dimensional line, only,
and a conjugation region is not maximally smooth. }. Thus, the
space-time is what the test particle can permanently visit due to
the work of engine\footnote{I do not see any `no-go' principle for
the existence of such the engine.}. An additional constraint is an
asymptotic boundary condition: a travel is starting from a region
of external observer well defined by the Schwarzschild solution at
large $r > r_g$.

\begin{figure}[th]
  \begin{center}
  \setlength{\unitlength}{1mm}
  \begin{picture}(140,70)
  \put(0,0){\includegraphics[width=15cm]{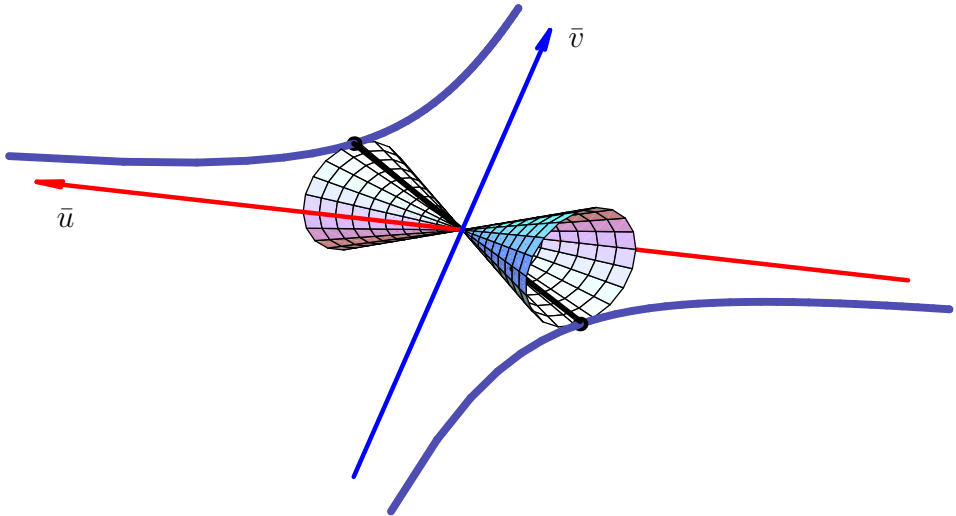}}
  \put(84,64){$\bar v$}
  \put(16,40){$\bar u$}
  \end{picture}
  \end{center}
  \caption{Physical space-times for geodesics of classes $CII$ (Kruskal)
  and $CIII$ (conic) with the common element $\bar u =\bar v$.
  }\label{pic2}
\end{figure}

So, geodesics completely confined under the horizon can be
described in terms of variables $\{\rho,\varphi_E\}$ introduced by
\begin{equation}\label{eucliddef}
    \left\{\begin{array}{l}
 \bar u=\,\varkappa\,{\rm i}\,\rho\,e^{{\rm i}\,\varphi_E},\\[2mm]
 \displaystyle
 \bar v=-\frac{{\rm i}}{\varkappa}\,\rho\,e^{-{\rm i}\,\varphi_E},\end{array}
 \right.
\end{equation}
or (at an `initial' time $\Delta t_0 =-2 r_g\ln\varkappa$, which
is set to zero)
\begin{equation}\label{euclid-def2}
    \left\{\begin{array}{l}
\;\,\, t=r_g\,\ln e^{-2{\rm i}\,\varphi_E}=-{\rm
 i}\,2r_g\,\varphi_E,\qquad \varphi_E\in [0,2\pi],
 \\[2mm]\displaystyle
 r_*= 2 r_g \ln\left[-\frac{\rho}{2 r_g}\right],
  \hfill\rho \in [0,2r_g],\end{array}
 \right.
\end{equation}
so that the interval takes the form
\begin{equation}\label{euclid-s}
    {\rm d}s^2=\frac{r_g}{r}\cdot e^{\displaystyle-\frac{r}{r_g}}\cdot({\rm
    d}\rho^2+\rho^2{\rm d}\varphi_E^2)-r^2({\rm d}\theta^2+\sin^2\theta{\rm
    d}\phi^2).
\end{equation}
Therefore, transition (\ref{eucliddef}) is periodic in $\varphi_E$
with the period $\Delta\varphi_E=2\pi$. Then, we get the period in
imaginary time
\begin{equation}\label{beta}
    \beta =4\pi r_g.
\end{equation}
We have to emphasize that transformation (\ref{euclid-def2}) is
not a formal Wick rotation \cite{York}, since it corresponds to
the space-time composed by causal geodesics continuously connected
with causal geodesics in Kruskal coordinates. Thus, from
(\ref{eucliddef}), (\ref{euclid-def2}) and (\ref{beta}) we deduce
that geodesics confined under the horizon should be periodic in
$\varphi_E$ or imaginary time, and they compose a statistical
ensemble with the Hawking--Gibbons temperature
\begin{equation}\label{T}
    T=\frac{1}{4\pi r_g}.
\end{equation}
The trajectory $\varphi_E=\pi/2$ reproduces $\bar u
=\varkappa^2\bar v$. Generically, the radial geodesics in range
$CIII$ propagate in the conic geometry of (\ref{euclid-s}) (see
Fig. \ref{pic2}), whereas $\rho=0\Leftrightarrow r=r_g$, and
$\rho=2 r_g\Leftrightarrow r=0$.

\begin{figure}[th]
\setlength{\unitlength}{1mm}
\begin{picture}(50,50)
\put(5,0){\includegraphics[width=40\unitlength,height=40\unitlength]{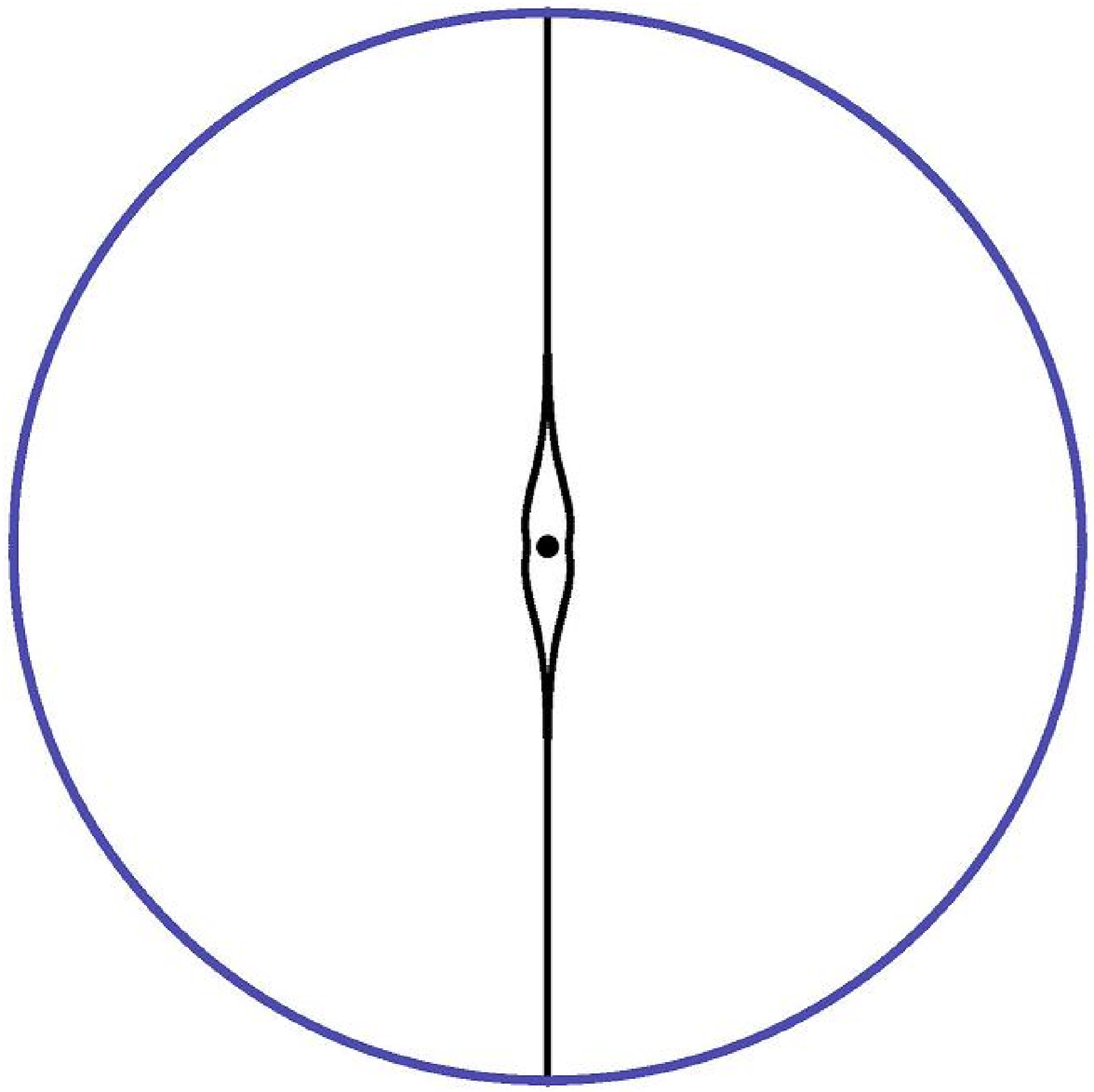}}
\put(55,0){\includegraphics[width=40\unitlength,height=40\unitlength]{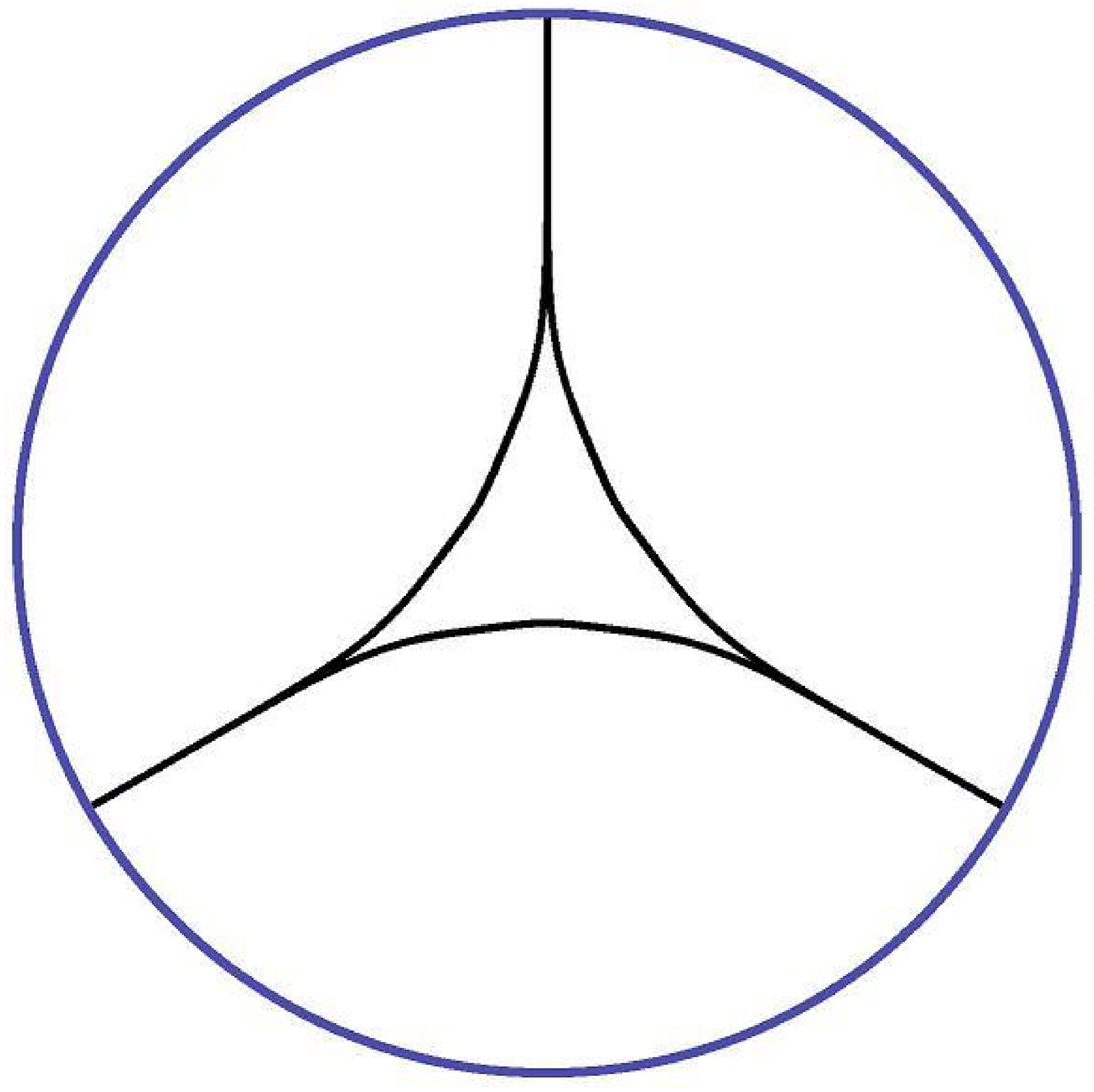}}
\put(105,-1){\includegraphics[width=40\unitlength,height=40\unitlength]{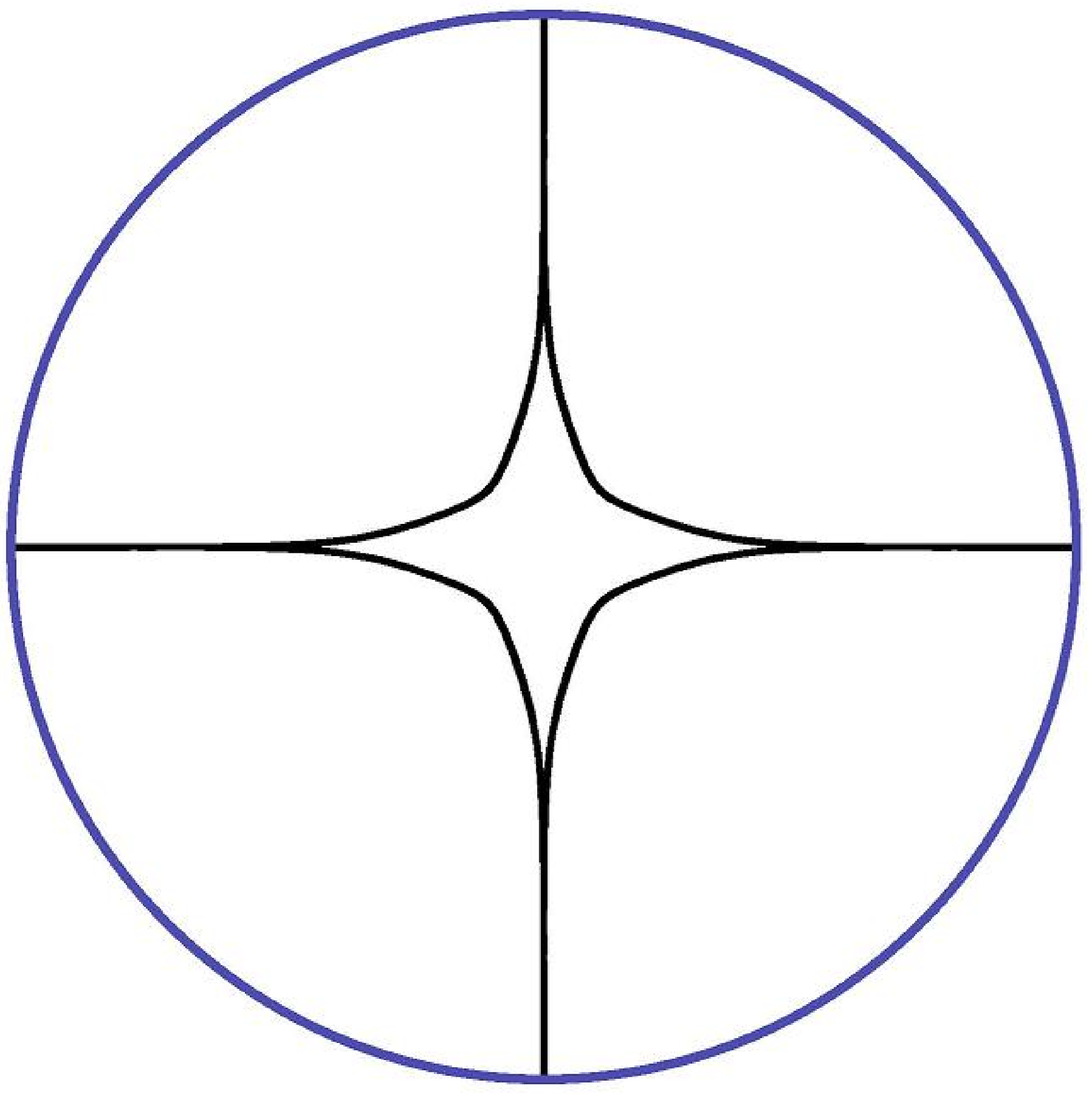}}
\end{picture}
\caption{The geodesics of $CIII$ with winding numbers 2, 3, 4 in
polar coordinates $\{\rho,\varphi_E\}$.} \label{pic3}
\end{figure}

For a complete description we deduce
\begin{equation}\label{derE}
    \frac{{\rm d}\varphi_E}{{\rm d}r} =
    \frac{1}{2r_g}\,\sqrt{1-\frac{r_c}{r_g}}\,\frac{r}{r_g-r}\,\sqrt{\frac{r}{r_c-r}},
\end{equation}
which allows us to calculate a cycle change of $\varphi_E$ with
$r$ varying from $0$ to $r_c$ and back:
\begin{equation}\label{cycle}
    \Delta_c\varphi_E = 2\int\limits_{0}^{r_c}{\rm d}r\,\frac{{\rm d}\varphi_E}{{\rm
    d}r}=
    \frac{\pi}{2}\,
    \bigg[2-(2+x)\sqrt{1-x}\bigg],
\end{equation}
where we have introduced $x=r_c/r_g$. The `winding' number (or
occupation number) is defined by a number of cycles in the period,
so that
\begin{equation}\label{wind}
    n = \frac{2\pi}{\Delta_c\varphi_E}.
\end{equation}
Since the motion should be periodic, we get the `quantization'
\begin{equation}\label{quant}
    n \in {\mathbb N}.
\end{equation}
Then, it results in the implicit quantization of $r_c$, i.e.
$x=x_n$. The occupation number monotonically grows with decreasing
of $x\in [0,1]$. The minimal $n$ is given by the limit $x\to 1$,
so that from (\ref{cycle}), (\ref{wind}) we deduce
\begin{equation}\label{min}
    n_{\rm min} =n_{x\to 1} =2.
\end{equation}
A picture of quantized geodesics is shown in Fig. \ref{pic3}.

\begin{figure}[th]
  \setlength{\unitlength}{1mm}
  \begin{center}
  \begin{picture}(110,70)
  \put(10,0){\includegraphics[width=10cm]{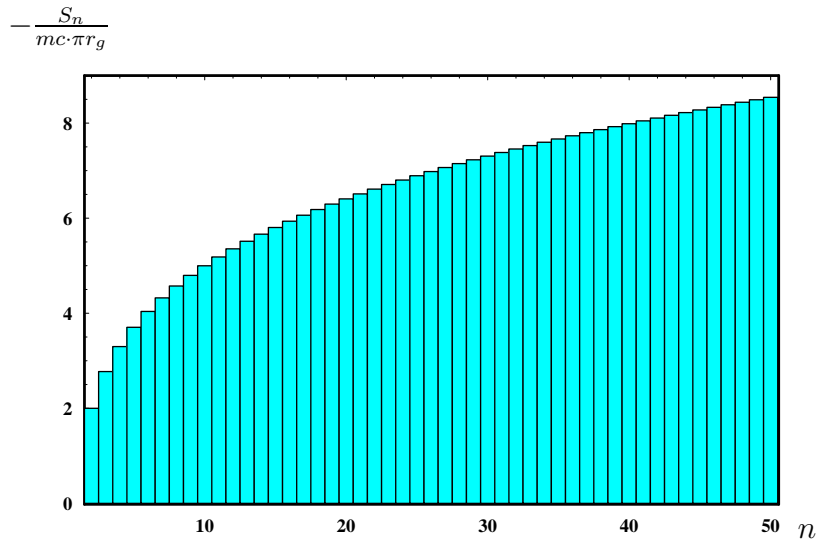}}
  \put(5,67){$-\frac{S_n}{mc\cdot \pi r_g}$}
  \put(110,0){$n$}
  \end{picture}
  \end{center}
  \caption{The quantum action.}\label{pic4}
\end{figure}

The action of relativistic particle
\begin{equation*}
    S=-mc\int{\rm d}s
\end{equation*}
at the geodesics confined under the horizon, is also quantized
\begin{equation}\label{quant-s}
    S_n =-mc\cdot n\int\limits_{{\scriptstyle cycle}}{\rm d}r\,\frac{{\rm d}s}{{\rm d}r}
    =-mc\cdot \pi r_g\,n\, \sqrt{x_n^3},
\end{equation}
as shown in Fig. \ref{pic4}.

The difference between two initial quantum actions grows with $M$
\begin{equation*}
    S_2-S_3 \approx  \frac{\pi}{c}\,G M m (3\cdot 0.924-2) \sim |S_2| \gg 1,\qquad
    M\to \infty,
\end{equation*}
at any nonzero mass of probe particle.

\begin{figure}[h]
\begin{tabular}{ccccccc}
\includegraphics[width=3cm]{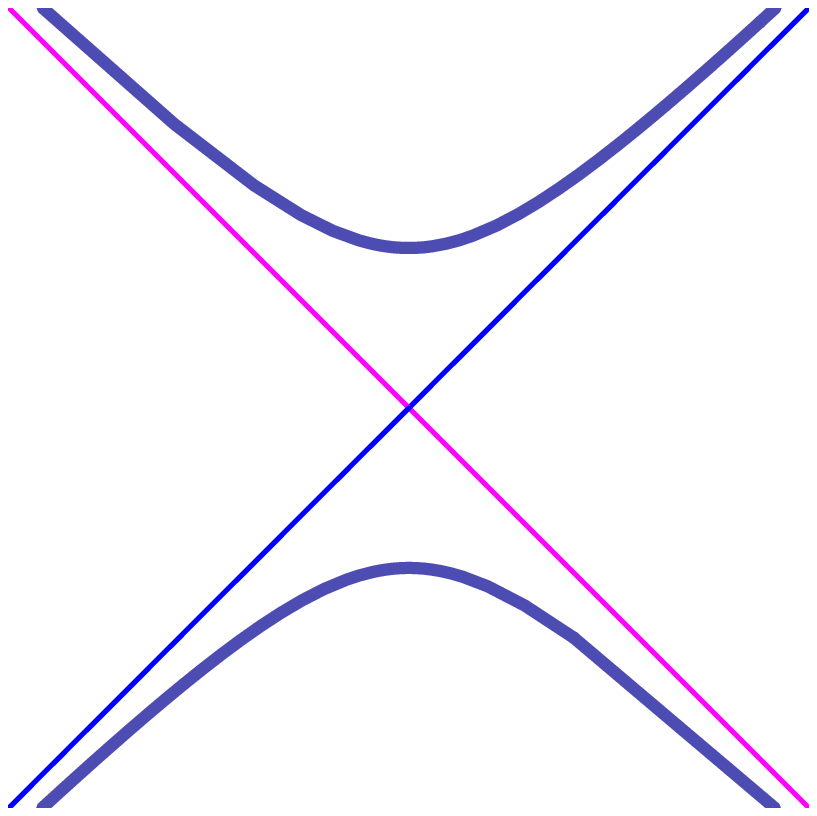} & \raisebox{1.35cm}{$\rightarrow$} &
\includegraphics[width=3cm]{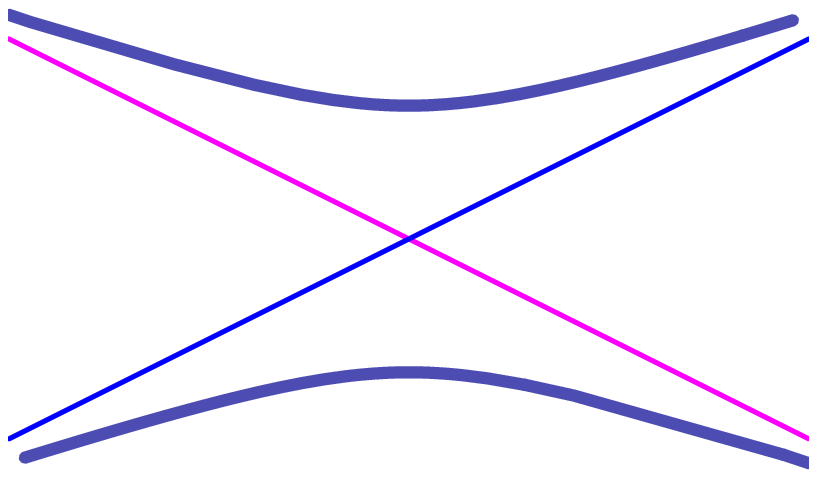} & \raisebox{1.35cm}{$\rightarrow$} &
\includegraphics[width=3cm]{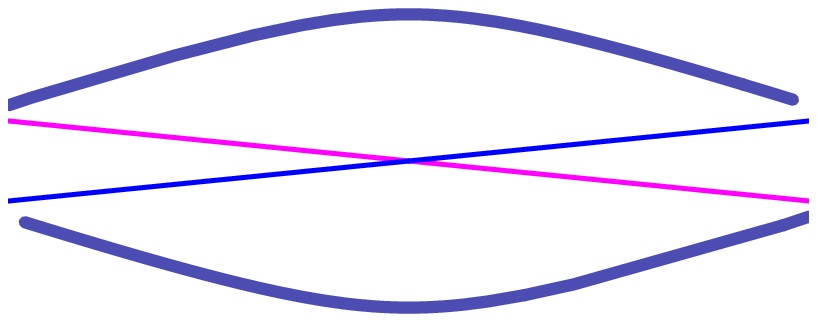} & \raisebox{1.35cm}{$\rightarrow$} &
\hspace*{-7mm}\includegraphics[width=3cm]{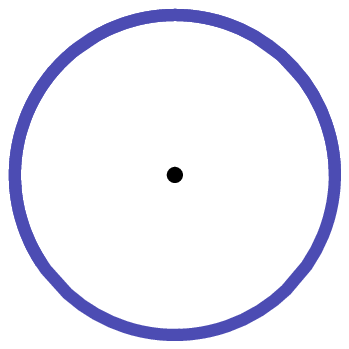}
\end{tabular}
\caption{Compactifying the hole.}\label{pic5}
\end{figure}

Next, in addition to quantization, class $CIII$ does not include
light cones (excluding a consideration of non-radial
motion)\footnote{In fact, the light on the horizon or, more
generally, massless particles can be prescribed to a trajectory
being a circle around the point $r=r_g$ with an infinitely small
radius. Then, the winding number $n=1$ corresponds to the massless
modes. We put the classical action of such modes equal to zero.}.
This fact helps us to imagine the $\{\rho,\varphi_E\}$ solid
circle as `a compactification' of holes in the $\{\bar u,\bar v\}$
Kruskal coordinates: one should glue the horizons and identify
them to a single point as shown in Fig. \ref{pic5}.

Finally, we have given the complete classification of radial
geodesics for the Schwarzschild black hole.

\section{Entropy}
As we have just seen, the geodesics confined under the horizon
periodically evolve with imaginary time. Therefore, they compose a
statistical ensemble with the period equal to the inverse
temperature $\beta$ (\ref{beta}). The partition function of
statistical ensemble is calculated as the evolution operator
summed over all trajectories with the given period. For
`classical' particles, the classical trajectories are taken into
account, only. For real time, the evolution operator is given by
\begin{equation*}
    U = \prod\sum e^{{\rm i} S},
\end{equation*}
where $S$ denotes the particle action, while the sum is running
over the particle trajectories, and the product is taken over
particles. The motion in imaginary time $t_E={\rm i}t$ gives
\begin{equation*}
    {\rm i} S ={\rm i}\int {\rm d}t\,L=\int{\rm d}t_E\,L =S_E,
\end{equation*}
where $L$ is a lagrangian, and $S_E$ is the `euclidian' action.
Then, the partition function is determined by
\begin{equation}\label{parti}
    Z = \prod\sum e^{S_E}.
\end{equation}
For the statistical ensemble of particles moving on geodesics
confined under the horizon, the euclidian action under study is
exactly the action we have just calculated.

For a given particle, as we have noted in the previous section, in
the limit of large black hole mass the splitting between the
action of winding number 2 and actions of higher winding numbers
is large, i.e. the action gap grows with $M$, so that the sum over
trajectories can be approximated by the leading term of $S_2$:
\begin{equation}\label{s2}
    \sum e^{S_E}\approx e^{S_2},
\end{equation}
whereas corrections are exponentially suppressed. Then, the
product over the particles in the partition function is reduced to
the sum of leading-term actions over particles,
\begin{equation*}
    Z\approx \prod e^{S_2} =e^{\sum S_2}.
\end{equation*}
The sum is easily calculated in the form
\begin{equation}\label{sum}
    \sum S_2 = -2\pi c\, r_g\sum m.
\end{equation}
Suppose that the gravitational field of black hole is generated by
the ensemble of particles completely confined under the horizon.
Then, it would be natural to suggest that the sum of particle
masses is equal to the mass of black hole\footnote{Actually, the
expression is fixed by a further evaluation of average energy (see
below).}:
\begin{equation}\label{mass}
    M=\sum m =\frac{1}{2}\,r_g,
\end{equation}
in the units $c^2=1$, $G=1$, $\hbar=1$. In these units, the sum of
actions is given by
\begin{equation}\label{s-e}
    \sum S_2 =-\pi\,r_g^2=-\frac{\beta^2}{16\pi},
\end{equation}
and the partition function is equal to
\begin{equation}\label{parti-2}
    Z = \exp\left[-\frac{\beta^2}{16\pi}\right].
\end{equation}

Thermodynamic relations give the following:
\begin{itemize}
    \item[1.] the average energy equal to the black hole mass,
    \begin{equation*}
    \langle E\rangle =-\frac{\partial\ln Z}{\partial\beta} =
    \frac{\beta}{8\pi} =\frac{r_g}{2} =M,
\end{equation*}
    \item[2.] the entropy
    \begin{equation*}
    {\cal S} = \beta\langle E\rangle+\ln Z
    =\frac{\beta^2}{16\pi} =\pi\,r_g^2,
\end{equation*}
    which is the Bekenstein--Hawking relation between the entropy
    and the area of horizon ${\cal A}=4\pi\,r_g^2$:
    \begin{equation*}
    {\cal S}=\frac{1}{4}{\cal A}.
\end{equation*}
\end{itemize}
Thus, in the leading approximation we have calculated the
partition function of static ensemble composed by particles at the
geodesics completely confined under the horizon of black hole.
Putting the sum of particle masses equal to the mass of black hole
we find the entropy of black hole at the microscopic level.

As for corrections to the partition function, they could be
twofold. First, terms by higher winding numbers $n$ are
exponentially suppressed, so that one can easily estimate a
convergency of series over $n$. Indeed, at large $n$ from
(\ref{cycle}), (\ref{wind}) and (\ref{quant-s}) we deduce
\begin{equation}\label{limS}
    S_n\to - mc\,\pi
    r_g\,\frac{16}{3}\,\sqrt[4]{\frac{3n}{16}},\qquad n\to \infty,
\end{equation}
so that terms of series \textit{exponentially} tend to zero with
$n$, and series is converging.

Second, a reasonable source of corrections to the partition
function is a non-radial motion, which we have not considered at
all, since we believe that for the black hole of zero angular
momentum the contribution of non-radial geodesics should be
suppressed.

\section{Discussion and conclusion}

\begin{figure}[th]
\centerline{\includegraphics[width=9cm]{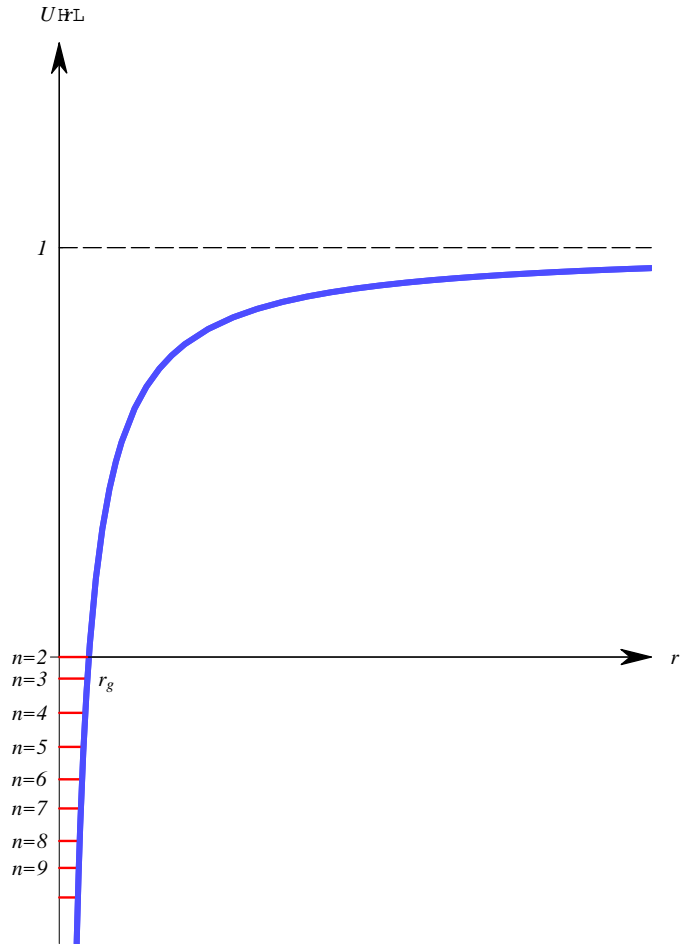}}
  \caption{The gravitational potential $U(r)$ and energy levels of
  particle confined under the horizon with various winding numbers.}\label{pic6}
\end{figure}

We have calculated the entropy of Schwarzschild black hole by
evaluating the action for radial geodesics completely confined
under the horizon, which form the statistical ensemble with the
Hawking temperature, in the leading order at large masses of black
holes. This procedure has become reasonable under the specific
quantization of trajectories by the winding number in the periodic
compact variable of conic geometry induced by the confined
geodesics. In this respect the Hamilton--Jacobi equation of
(\ref{3}) can be represented in the form
\begin{equation}\label{U}
    \frac{1}{m^2}\,\left(\frac{\partial{\cal S}_{HJ}}{\partial
    r_*}\right)^2 = {\cal E}_A-U(r),\qquad {\cal E}_A =
    \frac{1}{A},\quad U(r)=g_{tt}(r)=1-\frac{r_g}{r},
\end{equation}
which determines the spectrum of the problem, shown in Fig.
\ref{pic6}, where the region of ${\cal E}_A<0$ represents a
quantum thermal bath. Note, that all classical geodesics with
${\cal E}_A<0$ are permitted, however, a thermal equilibrium takes
place if only the trajectories are periodic in the euclidian time.

The rule of quantization
\begin{equation}\label{quant1}
    \oint {\rm d}\varphi_E=n\cdot\int\limits_{\Delta_c} {\rm d}\varphi_E =2\pi
\end{equation}
can be simply rewritten in the form
\begin{equation}\label{quantT}
    \oint T\,{\rm d}t_E =n\cdot\int\limits_{\Delta_c} T\,{\rm d}t_E={\hbar},
\end{equation}
where $n=2,3,\ldots$ is the winding number, $T$ is the
temperature, and $\Delta_c$ denotes the contribution per cycle.
One can represent (\ref{quantT}) as
\begin{equation}\label{quantTn}
    \oint T_n\,{\rm d}t_E =n\,{\hbar},\quad\Rightarrow\quad
    \oint {\rm d}T_n\,{\rm d}t_E ={\rm d}n\cdot{\hbar},
\end{equation}
where $T_n =n\,T$ is `the temperature associated to the $n$-th
level of winding number'. This rule can be compared with the
quasi-classical quantization by Bohr--Sommerfeld
\begin{equation}\label{BS}
    \oint p\;{\rm d}q =2\pi n\,\hbar,
\end{equation}
where $p$ is canonically conjugated to a dynamical variable $q$,
periodic in time.
\setcounter{footnote}{1}

Let us use the quasi-classics in order to count a number of
microstates ${\cal N}$ and to compare it with the entropy
evaluated by the partition function. So, in the case under study,
a role of evolution parameter is playing by the radius $r$ (or
$\rho$), while the euclidian time $t_E$ (or $\varphi_E$) is the
dynamical variable. It is simply to recognize that the
corresponding conjugated momentum is the euclidian energy ${\cal
E}_n =m_n/\sqrt{-A_n}$, where we have put index $n$ to the
particle mass to show the following: one could adiabatically vary
the masses of particles composing the thermal equilibrium, when
they occupy different levels with no change of
entropy\footnote{The action of particle can be rewritten in the
form $S=-mc\int{\rm d}r\,L_r$, at $L_r=\sqrt{g_{\mu\nu} {\rm
d}x^\mu/{\rm d}r\,{\rm d}x^\nu/{\rm d}r}$, so that the momentum
conjugated to $t_E$ is $p_E =-mc {\partial L_r}/{\partial( {\rm
d}t_E/{\rm d}r)}$, which is equal to $E$.}: for instance, if all
particles are situated at the $n$-th level, the calculated product
of interval per cycle $\Delta_c s_n$ by the sum of masses at
$n$-th level $\sum m_n$ and winding number should be invariant,
\begin{equation*}
    n\cdot\Delta_c s_n\,\sum m_n c =2\cdot\Delta_c s_2\,\sum
    m_2 c\equiv -S_2.
\end{equation*}
The invariance of action under such the re-distributions of
particles between the levels conserves the entropy. This variation
is responsible for the entropy as a measure of chaos. Then, using
the definition of temperature as an average energy per particle
\begin{equation*}
    T_n = \frac{E_n}{\nu},\qquad E_n=\sum {\cal E}_n,
\end{equation*}
where $\nu$ is a number of particles at the $n$-th level, `in a
quasi-classical manner' of (\ref{quantTn}) by (\ref{BS}) we get
\begin{equation}\label{eBS}
    \frac{1}{\nu}\oint E_n{\rm d}t_E =
    \hbar\,n,\qquad
    {\rm d}E_n\oint{\rm d}t_E =
    \hbar\,\nu\,{\rm d}n,
\end{equation}
Thermodynamically, we write down
\begin{equation*}
    {\rm d}E_n =T_n\,{\rm d}{\cal S} =n\; T\,{\rm d}{\cal S}
\end{equation*}
and deduce
\begin{equation}\label{dE2}
    T\,{\rm d}{\cal S}\oint
    {\rm d}t_E=
    \hbar\,\nu\,\frac{{\rm
    d}n}{n}=\hbar\,{\rm d}\ln{\cal N},
\end{equation}
where ${\cal N}=n^\nu$ is the total number of microstates for
$\nu$ particles each having $n$ states. Integrating (\ref{dE2})
over all levels, particles and their masses, we obtain
\begin{equation}\label{intE}
    \int T\,{\rm d}{\cal S}\oint
    {\rm d}t_E
    =
    \hbar\int\limits_2^{\sqrt[\nu]{{\cal N}}}\nu\,\frac{{\rm d}n}{n}\approx
    \hbar\,\ln{\cal N},\qquad {\cal N}^{\raisebox{1mm}{$\scriptstyle 1/\nu$}}\gg 1.
\end{equation}

On the other hand, the quantity
\begin{equation}\label{defI}
    {\cal I} =\int T\,{\rm d}{\cal S}\oint
    {\rm d}t_E = \hbar \int {\rm d}{\cal S} = \hbar\,{\cal S}
\end{equation}
is an adiabatic invariant value independent of $n$, i.e. the
entropy multiplied by $\hbar$.

Our calculations have given
\begin{equation}\label{I}
    {\cal I}=
    \hbar\,\frac{\pi r_g^2}{\ell^2_{\rm Pl}},
\end{equation}
where we have restored the fundamental units and denoted the
square of Planck length by $\ell^2_{\rm Pl}=G\hbar/c^3$.
Comparison of (\ref{defI}) with (\ref{intE}) gives the standard
correspondence between the entropy and number of microstates,
\begin{equation}\label{correspond}
    {\cal S} =\ln{\cal N}.
\end{equation}
Of course, we may treat the logical chain
(\ref{eBS})--(\ref{correspond}) backward, to derive the
quantization rule (\ref{eBS}) from the known expression for the
entropy.

The winding number is restricted by a maximal value, since
$\nu\geqslant 1$, and hence, the limit is $n_{\rm max}={\cal
N}=\exp{\cal S}\gg 1$. Let us remind that the action gap was
essential for the derivation of partition function, so that $M
m/m^2_{\rm Pl}\gg 1$, and the particle mass is restricted from
bottom. At $n\gg 1$ the appropriate mass tends to zero (see
(\ref{limS}))\footnote{Note also, that at $n\to n_{\rm max}$ the
energy of particle gets the value about the black hole mass
$E_{\cal N}=m_{\cal N}/\sqrt{-A_{\cal N}}\approx 3 M/8$.}, and it
should reach a minimal admissible value, so that $\nu\gg 1$.

Classical trajectories unsatisfying (\ref{quant1}) are not
forbidden, but they do not belong to a thermal bath with a
definite temperature, i.e. they are not in the thermal
equilibrium.

Such the quantization intrinsically differs from the
quasi-classics in the quantum mechanics, since in nature it
unavoidably involves the temperature: it seems that any real
theory of quantum gravity should by origin include the temperature
(or statistical chaos?). The exterior of black hole, i.e. the
space-time outside the horizon, forms the thermostats of thermal
geodesics confined under the horizon.

Having calculated the entropy by the partition function, we have
neglected the contributions by higher winding numbers (or used the
invariance of action by adjusting the sum of masses to get the
given mass of black hole) as well as by a non-radial motion.

The leading term with the minimal winding number gives the state,
which is `coherent' in a sense, that the geodesic, connecting
$r_g$ to $r=0$, matches all $r$ to the point $r=r_g$. Due to the
coherence, the physics under the horizon can be completely
described by the physics on the surface of the horizon. This fact
reveals the principle of holography: we do not need a three
dimensional interior, its surface is sufficient \cite{Holograph}.
The information inside the hole is equivalent to its snapshot of
horizon surface. Therefore, the calculation of entropy based on
the two dimensional conformal field theory (2D CFT) by the Cardy
formula for the number of states \cite{Cardy} is not in a
straightforwardly visible contradiction with the given
representation. Though we cannot find a direct connection between
our approach and the method based on 2D CFT, since we have not
introduced quantum fields. The same note concerns for the
comparison with the superstring calculation of entropy by counting
appropriate microstates (the vehicle is armed by the Cardy
formula, again).

The construction offered is, in a sense, dual to the Hawking
calculation based on the evaluation of gravitational action for
the given metric itself, since we deal with the action for the
particles moving in the metric. Therefore, we have implicitly
supposed that the metric is an effective average collective
gravitational field of particles. A duality is also suggestive in
connection to a method by Frolov and Novikov, who constructed a
calculation separating visible and invisible modes of quantum
field \cite{FrolovNovikov}. The former is radiated by the black
hole, while the later is eaten by the hole. In this sense, the
entropy is given by the information eaten by the black hole.

In conclusion, we have presented a picture for the interior of
Schwarzschild black hole, that composes the first part of serial
article, which will be continued by an analogous consideration of
Reissner--Nordstr\o m black hole.

\vspace*{5mm} The author expresses a gratitude to academician
A.A.Logunov, who has asked me for a description 
of radial geodesics in the metric of Schwarzschild black hole. I
specially thank associated professor O.I.Tolstikhin for a useful
remark.

This work is partially supported by the grant of the president of
Russian Federation for scientific schools NSc-1303.2003.2, and the
Russian Foundation for Basic Research, grants 04-02-17530,
05-02-16315.


\end{document}